\documentclass[prb,twocolumn,preprintnumbers,superscriptaddress]{revtex4-2}
\usepackage{graphicx}
\usepackage{bm}
\usepackage{amssymb}
\usepackage{amsmath}
\usepackage{color}
\usepackage{notes2bib}
\DeclareMathOperator{\e}{e}

\newcommand{\av}[1]{\langle #1 \rangle}
\newcommand{\fd}[1]{\text{D} [#1]}

\begin{document}

\title{Shot noise in resonant tunneling: Role of inelastic scattering}

\author{I.~V. Krainov}
\email{igor.kraynov@mail.ru} 
\affiliation{A.~F. Ioffe Physical-Technical
Institute of the Russian Academy of Sciences, 194021 St.~Petersburg,
Russia} 
\author{A.~P. Dmitriev}
\affiliation{A.~F. Ioffe Physical-Technical
	Institute of the Russian Academy of Sciences, 194021 St.~Petersburg,
	Russia}
\author{N.~S. Averkiev}
\affiliation{A.~F. Ioffe Physical-Technical
Institute of the Russian Academy of Sciences, 194021 St.~Petersburg,
Russia}

\date{\today}

\begin{abstract} 
We study the influence of inelastic processes on shot noise and the Fano factor for a one-dimensional double-barrier structure, where resonant tunneling takes place between two terminals. 
Most studies to date have found, by means of various approximate or phenomenological methods, that shot noise is insensitive to dephasing caused by inelastic scattering. 
In this paper, we explore the status of this statement by deriving a general Landaur-B\"uttiker-type formula that expresses the current noise and Fano factor in a one-dimensional conductor through inelastic scattering amplitudes.
For a double-barrier structure, exact scattering amplitudes are calculated in the presence of a time-dependent potential that acts in the region between the barriers. This allows us to rigorously analyse the role of dephasing in the current noise generated by applying a finite bias voltage to the resonant level. As an example of dephasing potential, we consider the one induced by equilibrium phonons. We show that for phonons propagating in one dimension, the random phase of the electron wave function, which is induced by the electron-phonon coupling, exhibits a diffusion-like dynamics. At the same time, for higher-dimensional phonons, the electron phase dynamics turns out to be non-diffusive, such that the average square of the phase grows logarithmically with time.
We calculate transmission coefficients of a double-barrier structure for these two types of phonon-induced dephasing. In the case of diffusive phase relaxation, the resonant level has a Lorentzian shape with the broadening determined by a sum of the elastic linewidth and the phase breaking rate. Logarithmic dephasing leads to an unusual shape of the size-quantized level: the transmission coefficient is characterized by the two energy scales, one governed by the transparency of barriers and the other by the phonon correlation time. We further calculate the Fano factor for these types of dephasing, using exact expressions for inelastic transmission and reflection amplitudes. 
It turned out that when an integer number of levels fall into the energy window of width eV, where V is the voltage applied to the structure, the Fano factor is really insensitive to inelastic processes inside the structure and coincides with the prediction of phenomenological models with an accuracy of small corrections depending on these processes.  On the contrary, at low voltages, when the eV window is smaller than the level width, this dependence is particularly pronounced and the phenomenological formula does not work. 
\end{abstract} 

\maketitle

Phase coherence of electron waves plays an essential role in low-dimensional transport \cite{Review}. Inelastic electron scattering processes, such as electron-electron or electron-phonon collisions, break down the phase coherence, leading to dephasing of electron waves. 
With decreasing intensity of such collisions (e.g., with lowering temperature), an electron system 
exhibits a crossover from the classical regime to the quantum one, where interference effects become prominent.

One of the simplest and important for practice interference phenomena is the so-called resonant tunneling through a quantum dot with tunnel contacts. At very low temperatures, the role of inelastic collisions is negligible and the conductance of the dot demonstrates a narrow high peak whenever the Fermi level coincides with one of the size-quantization levels. The appearance of these conductance peaks is due to interference of electron waves inside the quantum dot. The width of the resonant peaks is determined by the transparency of the contacts that represent tunneling barriers for electrons. With increasing temperature, the interference breaks down because of the inelastic collisions, the peaks widen and gradually disappear. When the phase coherence is completely destroyed, the conductance of the double-barrier structure is given by a classical formula containing only the probabilities of tunneling through the contacts. 

This phenomenon of resonant tunneling was studied experimentally in various systems, such as carbon nanotubes \cite{CNT_exp_PRB_2005,CNT_exp_nature,CNT_exp_nano_lett,CNT_exp_book,ACSNano2017} and nanowires with tunnel contacts, bulk structures with double barriers \cite{QB_exp,Esaki,DBRTD_PRB_SE,QW_DBRTD_sse,DBRTD_PRB_SE_Science}, and many others.
Theoretically, the effect of inelastic scattering on resonant tunneling was considered in detail in Ref.~\cite{Lee_PRL}.
It was shown there, in particular, that although the width of the resonant levels is determined both by the transparency of the barriers and by the rate of inelastic processes, the conductance at temperatures exceeding this width is determined solely by the transparency of the barriers. 

Another important characteristic of such structures is the current-noise power \cite{BLANTER20001,Kobayashi2021}. Current noise consists of two terms -- thermal and non-equilibrium noise.
The Landauer-B\"uttiker scattering formalism, considering a system attached to Fermi leads and characterizing the structure by the scattering amplitudes, is a common tool for the analysis of the current noise. 
The thermal (or Johnson-Nyquist) noise is proportional to the product of the temperature $T$ and the conductance $G$ of the structure (proportional to its transparency): $$ S_T(\omega = 0) = 2 T G.$$
The non-equilibrium noise at zero temperature (shot noise) is proportional to the product of the transmission coefficient 
$\mathcal{T}$ and the reflection coefficient $\mathcal{R}=1-\mathcal{T}$
of the scatterer:
$$ S \propto \mathcal{T}(1-\mathcal{T}).$$ 
The convenient measure of shot noise is the Fano factor -- the ratio of zero-frequency noise power to the Poissonian noise: 
\begin{equation}
  F = S(\omega = 0) / S_P.
  \label{eq1}
\end{equation}
The Poissonian noise is proportional to the average current: 
\begin{equation}
    S_P = e \langle I \rangle
    \label{eq2}
\end{equation}
(here, we considered spinless particles with charge $e$).

For resonant tunnelling through a double-barrier structure in the nonlinear transport regime (the temperature is smaller than the applied bias voltage), the Fano factor is
given by \cite{BLANTER20001}:
\begin{gather}
\label{FanoSimple}
    F = \frac{\Gamma_L^2 + \Gamma_R^2}{(\Gamma_L + \Gamma_R)^2}
    = \frac{\mathcal{T}_L^2 + \mathcal{T}_R^2}{(\mathcal{T}_L + \mathcal{T}_R)^2}.
\end{gather}
Here, $\Gamma_{L}\propto \mathcal{T}_L$ and $\Gamma_R\propto \mathcal{T}_R$ quantify the strength of the left and right barriers in energy units
The width of the resonant level in the absence of inelastic processes is given by $\Gamma_L + \Gamma_R$:
the stronger the barrier, the smaller its contribution to the level width.
The Fano factor given by Eq.~(\ref{FanoSimple}) can take values between 1/2 and 1. 
Expression (\ref{FanoSimple}) was obtained in Ref.~\cite{ChenF} and \cite{BUTTIKER1991199} using complementary approaches.

Remarkably, when written in terms of the transmission coefficients $\mathcal{T}_{L,R}$, the expression for the Fano factor can be interpreted both quantum-mechanically and classically.
Thus, an interesting feature of the resonant double-barrier structure is  apparent insensitivity of the Fano factor to inelastic processes in the system. This statement has been widely discussed in literature and tested by various methods. The classical tunneling picture based on master equation (describing fully incoherent transport) employed in Refs.~\cite{Wilkins,ChenNoise} or the approach based on the Langevin equation \cite{BLANTER20001} led to Eq.~(\ref{FanoSimple}). Within the quantum coherent picture of  transport, a solution of quantum master equation \cite{MilburnPRB} and calculations based on non-equilibrium Green functions \cite{Runge,Galperin} yielded the same result (\ref{FanoSimple}) for the Fano factor. 

Since in both the limits of fully coherent and fully incoherent transport, the Fano factor is given by the same formula, it is tempting to argue that coherence and its breaking by inelastic processes does not play any role here. However, this does not, of course, prove that, in the intermediate case of finite dephasing, the result remains unchanged. Indeed, considering all the interference-induced terms separately, the one can imagine the situation when this coherent contribution is exactly zero (complete destructive interference). Any finite dephasing, destroying such a cancellation, would lead to corrections to Eq.~(\ref{FanoSimple}), whereas extremely strong dephasing would kill all the interference terms, thus restoring Eq.~(\ref{FanoSimple}) in the classical limit.   

Several works indeed found non-universality of the Fano factor for the double-barrier structure.
In particular, Ref.~\cite{Iannaccone}
found corrections to this expression within the sequential-tunneling picture.
Another work, Ref.~\cite{WilkinsPhase},  demonstrated strong deviation from result (\ref{FanoSimple}) by  phenomenologically adding random (completely uncorrelated) phases to the quantum amplitudes that describe electron propagation  between the barriers. However, importantly, these papers used various assumptions and simplifications in the consideration of inelastic scattering. Thus, the question of influence of inelastic scattering on the Fano factor for resonant tunneling is still open. This calls for the analysis of inelastic processes within an exact quantum-mechanical model. The main goal of this paper is to develop such a formalism and to finally resolve the question about the influence of inelastic processes on the Fano factor of a double-barrier structure.



In this paper, we obtain a general expression for the noise power for transport through a one-dimensional scatterer in the presence of inelastic processes. The current noise is expressed in terms of
inelastic transmission and reflection amplitudes, which allows us to take dephasing into account exactly. 
Next, we obtain the inelastic transmission amplitudes for a double-barrier structure with random time-dependent potential acting between the barriers. Within this scattering formalism, we calculate shot noise for resonant tunneling in the presence of inelastic scattering. After averaging over the random potential, we obtain
the structure's Fano factor and analyze it for different types of dephasing of the electron wave function. Specifically, we investigate phonons 
as a source of the random potential leading to dephasing. If dephasing caused by one-dimensional (1D) phonons, the dispersion of the random wave-function phase $\varphi_f$ possesses a diffusion growth with time $t$, characterized by the dephasing time $\tau_\varphi$: 
$\langle \varphi^2_f \rangle \sim t / \tau_\varphi$. Remarkably, for two-dimensional (2D) and three-dimensional (3D) phonons, 
dephasing is logarithmic: 
$\langle \varphi_f^2 \rangle \sim \ln(Tt/\hbar)$. 

In the case of diffusive phase dynamics, the resonant levels in a double-barrier structure have a Lorentzian shape with the broadening determined by a sum of the elastic linewidth and the phase breaking rate. Logarithmic dephasing leads to an unusual shape of the size-quantized level: the transmission coefficient is characterized by the two energy scales, one governed by the transparency of barriers and the other by the phonon correlation time. 
When the voltage exceeds the characteristic resonant level width, the leading contribution to the Fano factor,  in agreement with the predictions of phenomenological models, becomes insensitive to the inelastic scattering, 
the influence of inelastic processes is only reflected in the presence of small corrections to this result. On the contrary, when the voltage is smaller than the level width, the Fano factor depends strongly on the type of dephasing and the phenomenological formula does not work.

\section{Current noise and Fano factor}
\label{Sec:I}

\begin{figure}[t]
	\centering
	\includegraphics[width=0.75\linewidth]{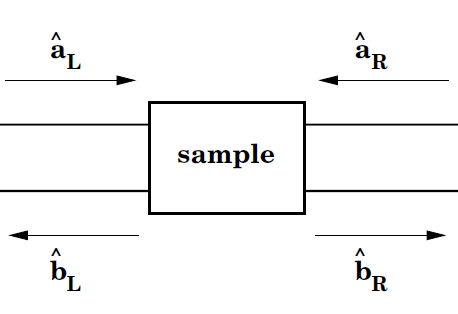}
	\caption{Schematics of the structure. The four electron operators of the right- and left-moving plain waves (shown by arrows) on both sides of the sample are related by the scattering matrix, Eq.~(\ref{sPowerDef}). There are only two independent operators, $\hat{a}_L$ and $\hat{a}_R$ supplied by the terminals; the other two, $\hat{b}_L$ and $\hat{b}_R$, are expressed through them by means of scattering amplitudes.}
	\label{structure}
\end{figure}

We start by deriving general formulas for the current in the 1D geometry within the scattering theory approach. In this section, we obtain the noise power in a 1D conductor with an inelastic scatterer, see Fig.~\ref{structure}. 
The current noise is determined by  
the current fluctuations:
\begin{gather}
\label{Noisedef}
S(t-t',x,x')=\frac{1}{2}\left\langle\! \delta \hat{I}(x,t) \delta \hat{I}(x',t') + \delta \hat{I}(x',t') \delta \hat{I}(x,t)\! \right\rangle,
\end{gather}  
where
\begin{align}
    \delta \hat{I}(x,t) = \hat{I}(x,t) - \langle \hat{I}(x,t) \rangle,
\label{devIdef}
\end{align}
$\hat{I}$ is the current operator, which takes the form 
\begin{gather}
\label{Idef}
\hat{I}(x,t) = \frac{\hbar e}{2 m i}
\left[ \hat{\Psi}^\dag (x,t) \frac{\partial \hat{\Psi} (x,t)}{\partial x}  -  \frac{\partial \hat{\Psi}^\dag (x,t)}{\partial x}   \hat{\Psi} (x,t)\right]
\end{gather}
for a single parabolic band characterized by the mass $m$,
$\hat \Psi$ is the electron field operator,
and $\langle\ldots\rangle$ denotes the expectation value determined by the statistical operator.

On the left of the scatterer, the wave-function operator for free waves is written as
\begin{gather}
\label{PsiDef}
\hat{\Psi}_L (x,t) = \int_0^\infty 
\frac{d k}{\sqrt{2 \pi}} \e^{-i E_k t} 
\left[ \e^{i k x} \hat{a}_L (k) + \e^{-i k x} \hat{b}_L (k) \right],
\end{gather}
where $\hat{a}_L (k)$ and $\hat{b}_L (k)$ 
are fermionic annihilation operators 
of the right- and left-moving waves with wave vector $k$ and energy $E_k$ in the left (hence index ``L'') part of the structure. Similarly, we define the the fermion field operator $\hat{\Psi}_L (x,t)$
to the right of the ``sample'' through the operators $\hat{a}_R (k)$ and $\hat{b}_R (k)$
describing the right- and left-moving waves, respectively, in the right (hence ``R'') part of the setup. 

In order to obtain the expectation value of the current, 
we need to express the field operators through the independent operators corresponding to the incident waves, $\hat{a}_L (k)$ and $\hat{a}_R (k)$, supplied by the two terminals. 
The scattered waves are expressed through those incident waves by means of the scattering matrix describing the sample (see Fig.~\ref{structure}).
Below, we will first consider the case of elastic scattering, and then generalize the scattering approach to include inelastic processes.

\subsection{Elastic scatterer}

In order to set the stage and introduce the formalism, we first analyze the case of elastic scattering and reproduce the known results for the current noise.  
In this case, the scattering matrix connects the incident and outgoing waves with the same wave-vector and is given by
\begin{gather}
\label{Sdef}
\left( \begin{matrix}
\hat{b}_L (k) \\
\hat{b}_R (k)
\end{matrix} \right) = 
\left( \begin{matrix}
\texttt{r}_L(k) && \texttt{t}(k) \\
\texttt{t} (k) && \texttt{r}_R(k)
\end{matrix} \right) 
\left( \begin{matrix}
\hat{a}_L (k) \\
\hat{a}_R (k)
\end{matrix} \right).
\end{gather}  
Here, the amplitudes $\texttt{r}_R(k)$ and $\texttt{r}_L(k)$ describe the reflection of the waves arriving from the right and left terminals, respectively. In the general case, these complex amplitudes differ by the phase, leading to the same reflection coefficient $\mathcal{R}=|\texttt{r}_R|^2=|\texttt{r}_L|^2$.
The transmission amplitudes $\texttt{t}(k)$ 
for the scattering of waves on both sides of the sample are identical when the time-reversal symmetry is preserved.  
The unitarity of the scattering matrix requires
\begin{align}
    \mathcal{T}+\mathcal{R}=1,
    \quad \texttt{r}_R\texttt{t}^*+\texttt{r}_L^*\texttt{t}=0, 
\end{align}
where $\mathcal{T}=|\texttt{t}|^2$ is the elastic transmission coefficient.
The scattering amplitudes for the elastic scatterer may depend on the wave-vector $k$ labelling the incoming waves (the same $k$ labels the outgoing waves). In what follows, for simplicity, we have assumed here that the right and left reflection amplitudes have the same phase and omit their subscripts (R/L).
In fact, the phases of reflection amplitudes are of no importance for the averaged current and noise in the elastic case. We will return to this point in the case of inelastic scattering below.

Using scattering matrix (\ref{Sdef}), we can write the current operator to the left of the scatterer through the operators $\hat{a}_L (k)$ and $\hat{a}_R (k)$
that describe the left and right terminals (Fig. \ref{structure}), respectively:
\begin{align}
\label{Iel}
&\hat{I}(x,t) = \frac{\hbar e}{4 \pi m} \int dk\, dk' \, \e^{i(E_{k'} - E_k)t} \\ 
&\times\! \bigg[ 
\hat{a}^\dag_L (k') \hat{a}_L (k) C_{LL}(k',k;x)\!+\!
\hat{a}^\dag_R (k') \hat{a}_R (k) C_{RR}(k',k;x) 
\notag
\\ 
&+\hat{a}^\dag_L (k') \hat{a}_R (k) C_{LR}(k',k;x)\!+\!  
\hat{a}^\dag_R (k') \hat{a}_L (k) C_{RL}(k',k;x) \bigg].
\notag
\end{align}
Here, the coefficients $C_{ij}(k,p;x)$ are 
expressed in terms of the scattering amplitudes as follows:
\begin{align}
\nonumber
C_{LL}&\! =\! (k+k') \left[\e^{i(k-k')x} - \texttt{r}^*(k')\texttt{r}(k) \e^{-i(k-k')x}\right]
\\ 
&-(k-k') \left[\texttt{r}^*(k')\e^{i(k+k')x} - \texttt{r}(k) \e^{-i(k+k')x}\right],
\\
C_{RR} 
&\!=\! - (k+k') \e^{i(k'-k)x} \texttt{t}^*(k')\texttt{t}(k), 
\\ 
\notag
C_{LR} 
&\!=\! - (k+k') \e^{i(k'-k)x} \texttt{r}^*(k')\texttt{t}(k) 
\\ 
&\quad
+ (k-k') \e^{-i(k'+k)x} \texttt{t}(k), 
\\ 
\nonumber
C_{RL} 
&\!=\! - (k+k') \e^{i(k'-k)x} \texttt{t}^*(k')\texttt{r}(k)  
\\ 
&\quad +(k'-k) \e^{-i(k'+k)x} \texttt{t}^*(k').
\end{align}

The current operator is represented as a double integral over the wave vectors.
Averaging over the states emanating from the left and right leads (independent terminals, $i,j=L,R$) produces, 
\begin{gather}
    \langle \hat{a}^\dag_i (k) \hat{a}_j (p) \rangle = \delta_{ij} \delta(k-p) f_i(k),
    \label{average-f}
\end{gather}
where $f_{i}$ is the electron distribution 
function of terminal $i$. The delta-function in Eq.~(\ref{average-f}) removes one of the integrals over the wave vectors in Eq.~(\ref{Iel}). 
In what follows, we will focus on the case of equilibrium terminals described by the Fermi-Dirac distribution.
Using Eq.~(\ref{average-f}),  
one obtains the average current,
which is now a single integral over the energy $E$ of the scattering states:
\begin{gather}
\langle \hat{I}(x,t) \rangle 
\label{Iavg}
= \frac{e}{h} \int dE\, \mathcal{T}(E) \left[f_L(E) - f_R(E)\right].
\end{gather}

Next, we turn to the calculation of the current-noise power. The statistical average of the fourth power of creation/annihilation operators
reads:
\begin{gather}
\label{a4def}
\langle \hat{a}^\dag_i \hat{a}_j \hat{a}^\dag_n \hat{a}_m \rangle  - \langle \hat{a}^\dag_i \hat{a}_j \rangle \langle \hat{a}^\dag_n \hat{a}_m \rangle = \delta_{im} \delta_{jn} f_m(1-f_n),
\end{gather}
where, for brevity, we denote
by a single index the wave vector and the terminal. The averaged correlation of the current fluctuations can be then written as
\begin{align}
\nonumber
\Big\langle \delta \hat{I}(x,t) \delta \hat{I}(x',t')\Big\rangle 
=& 
\left(\frac{\hbar e}{4 \pi m}\right)^2 \! \int dk\,dp\, \e^{-i(E_k - E_p)(t-t')} 
\\ 
\nonumber
\times\bigg[ f_L(p) [1-f_L(k)]&\, C_{LL} (p,k;x) C_{LL}(k,p;x')  
\\
\nonumber
+
f_R(p) [1-f_R(k)]&\, C_{RR}(p,k;x) C_{RR}(k,p;x') 
\\ 
\nonumber
+
f_L(p) [1-f_R(k)]&\, C_{LR}(p,k;x) C_{RL}(k,p;x') 
\\
+
f_R(p) [1-f_L(k)]&\, C_{RL}(p,k;x) C_{LR}(k,p;x') \bigg].
\label{curr-corr-elast}
\end{align}

The noise power for \textit{elastic scattering}, i.e., the zero-frequency noise  \cite{Beenaker}, is calculated from the current correlation function (\ref{curr-corr-elast}) with both current fluctuations are taken on one side of the scatterer:
\begin{gather}
\nonumber
S = \frac{e^2}{h} \int dE\, 
\bigg\{ 
\mathcal{T}(E) \left[ f_L(1-f_L) + f_R(1-f_R) \right]  \\
+ \mathcal{T}(E)[1-\mathcal{T}(E)] \left(f_L - f_R\right)^2 
\bigg\}.
\label{sElGeneral}
\end{gather}
In this formula, the first term describes thermal noise. In the limit of zero applied bias voltage,
the fluctuations are proportional to the temperature multiplied by the susceptibility -- the conductance in this case, in accordance with the fluctuation-dissipation theorem.  The second term is the shot noise, which is sensitive to the energy dependence of the transmission coefficient. 
If temperature is zero, only the shot-noise contribution is nonzero and one gets:
\begin{gather}
S = \frac{e^2}{h} \int_{\mu}^{\mu + eV} dE\, \mathcal{T}(E)[1-\mathcal{T}(E)],
\label{zero-T-elastic-shot}
\end{gather}
where the energy integral is restricted to the window determined by the bias voltage $V$ and $\mu$ is the chemical potential of one of the terminals.
Shot noise is quantitatively characterized by a single number -- the Fano factor, Eq.~(\ref{eq1}): 
\begin{gather}
\label{FanoFactorDefElastic}
F = \frac{\int dE\, \mathcal{T}(E)[1-\mathcal{T}(E)]}{\int dE\, \mathcal{T}(E)},
\end{gather}
where the integration is performed over the energy difference between the left and right terminals.

\subsection{Inelastic scatterer} 

Let us now turn to the case of an inelastic scatterer.
In order to take into account inelastic processes (caused, for example, by the electron-phonon interaction), we consider a general form of the scattering matrix:
\begin{gather}
\label{scatMatrInel1}
\hat{b}_L (k) = \int dk' \left[ \texttt{r}_L(k,k') \hat{a}_L (k') + \texttt{t}_{L}(k,k') \hat{a}_R(k') \right], \\ \label{scatMatrInel2}
\hat{b}_R (k) = \int dk' \left[ \texttt{t}_{R}(k,k') \hat{a}_L (k') + \texttt{r}_R(k,k') \hat{a}_R(k') \right]. 
\end{gather}
Here, the scattering amplitudes, which are now functions of the two wave vectors, are labelled by the indexes corresponding to the outgoing waves. 
Dependence on two wave vectors will be in the case of time dependent potential of scatter, which we calculate in next section. In this case inelastic processes break down the time-reversal symmetry. 
Hence, we introduce distinct transmission amplitudes for
the waves scattered to the right and to the left. Scattering amplitudes satisfy the unitary relations, which are now of the integral form: 
\begin{gather}
    \delta(k-k') = \int dp\, \left[ \texttt{r}_L^*(k,p) \texttt{r}_L(k',p) + \texttt{t}_L^*(k,p) \texttt{t}_L(k',p) \right],
    \label{unitarity-inelastic}
\end{gather}
and the same with $R\leftrightarrow L$.

The current operator to the left of the scatterer can be written as:
\begin{align}
\nonumber
\hat{I}(x,t) &= \frac{\hbar e}{4 \pi m} \int dkdk'dpdp' \e^{i(E_{k'} - E_k)t} \\ \nonumber
&\times\! \big[ 
\hat{a}^\dag_L (p) \hat{a}_L (p') 
C_{LL}\left(k,k';p,p';x\right)
\\
&\ +  
\hat{a}^\dag_R (p) \hat{a}_R (p') C_{RR}\left(k,k';p,p';x\right) 
\notag\\
&\ +  
\hat{a}^\dag_L (p) \hat{a}_R (p') C_{LR}\left(k,k';p,p';x\right)
\notag
\\
&\ +  
\hat{a}^\dag_R (p) \hat{a}_L (p') C_{RL}\left(k,k';p,p';x\right) \big]
\label{IcIn}, 
\end{align}
where the coefficients $C_{ij}\left(k,k';p,p';x\right)$ 
read
\begin{align}	
C_{LL} &=  
(k+k') 
\big[ \e^{i(k-k')x} \delta(k'-p)\, \delta(k-p')
\notag
\\
\notag
&- 
 \e^{-i(k-k')x} \texttt{r}^*_L(k',p)\, \texttt{r}_L(k,p') \big] 
 \\ 
 \nonumber
&+(k'-k) \big[ \e^{-i(k+k')x} \delta(k'-p)\, \texttt{r}_L(k,p') 
\label{CLL}
\\
&- \e^{i(k+k')x} \delta(k-p')\, \texttt{r}^*_L(k',p) \big], \\ 
C_{RR} &= - (k+k') \e^{-i(k-k')x} \texttt{t}^*_L(k',p)\, \texttt{t}_L(k,p'), \\ \nonumber 
C_{LR} &= -(k+k') \e^{-i(k-k')x} \texttt{r}^*_L(k',p)\, \texttt{t}_L(k,p')  \\ 
&+(k'-k) \e^{-i(k+k')x} \delta(k'-p)\, \texttt{t}_L(k,p'), \\ \nonumber
C_{RL} &= -(k+k') \e^{-i(k-k')x}\texttt{r}_L(k,p') \, \texttt{t}^*_L(k',p)  \\ 
&-(k'-k) \e^{i(k+k')x} \delta(k-p')\ \texttt{t}^*_L(k',p).
\label{CRL}
\end{align}
For the current to the right of the scatterer, one obtains an analogous expression with $R\leftrightarrow L$. After the statistical averaging, the current in the left lead takes a form:
\begin{align}
\label{av-I-inel}
& \langle \hat{I}(x,t) \rangle = \frac{\hbar e}{4 \pi m} \int dk\,dk'\,dp\, \e^{i(E_{k'} - E_k)t} \\ 
   & \times\bigg[ f_L(p) C_{LL}\left(k,k';p,p;x\right) + f_R(p) C_{RR}\left(k,k';p,p;x\right) \bigg].
\nonumber
\end{align}

Note that, in the presence of inelastic scattering, the expectation value of the current depends on time. 
In real experiments, in order to obtain the average current, one also performs the time averaging. We assume that this time averaging is done with a function $g_T(t)$ defining a wide time window $\tau_\text{av}$ -- the characteristic time of averaging. 
This time is the largest timescale in the problem. The window function $g_T(t)$ has the following properties:
\begin{gather}
    \int_{-\infty}^\infty dt\, g_T(t) = 1,\\ \label{gDef}
    \tau_\text{av} \int_{-\infty}^\infty dt\, g_T(t) \e^{i(E_k-E_{k'})t} \sim  \delta(E_k-E_{k'}). \\ \label{gIntDelta}
\end{gather}
We will denote the averaging with $g_T(t)$ as $\langle \ldots \rangle_t$. With the help of this averaging,
a useful relation can be obtained from the flow conservation for each incident $p$:
\begin{align}
&\int dk\,dk'\, \langle \e^{i(E_k-E_{k'})t} \rangle_t
\nonumber 
\\
&\times \bigg[ C_{LL}\left(k,k';p,p;x\right) + C_{RR}\left(k,k';p,p;x\right) \bigg] = 0.
\label{unitaryAverage}
\end{align}
Let us now define time-dependent scattering amplitudes ($j=R,L$):
\begin{gather}
    \int dk\, \e^{-i E_{k} t} \texttt{t}_j(k,p) = \tilde{\texttt{t}}_j(t,p),
    \label{ttp}
    \\
    \int dk\, \e^{-i E_{k} t} \texttt{r}_j(k,p) = \tilde{\texttt{r}}_j(t,p).
    \label{rtp}
\end{gather}
The flow conservation (\ref{unitaryAverage}) can be then cast into the following form [cf. Eq.~(\ref{unitarity-inelastic})]:
\begin{gather}
    1  = \langle \tilde{\texttt{r}}^*_L(t,p) \tilde{\texttt{r}}_L(t,p) \rangle_t + 
    \langle \tilde{\texttt{t}}^*_L(t,p) \tilde{\texttt{t}}_L(t,p) \rangle_t.
\end{gather}
Using this, we can find average current:
\begin{gather} \nonumber
    \Big\langle\langle \hat{I}(x,t) \rangle \Big\rangle_{t} = \frac{e}{h} \int dp\, v(p)\,
    \left\langle \tilde{\texttt{t}}^*_L(t,p) \tilde{\texttt{t}}_L(t,p) \right\rangle_t\, \\
    \cdot \Big[ f_L(p) - f_R(p) \Big],
\end{gather}
where $v(p)$ is the velocity at momentum $p$.
From now on, we will assume, for simplicity, a linearized dispersion relation for electrons, so that $v$ will not depend on $p$. 
The transmission coefficient in the inelastic case does not depend on whether the waves are supplied by the left terminal or by the right one:
\begin{gather}
    \mathcal{T} (p) = \left\langle \tilde{\texttt{t}}^*_R(t,p) \tilde{\texttt{t}}_R(t,p) \right\rangle_t = 
    \left\langle \tilde{\texttt{t}}^*_L(t,p) \tilde{\texttt{t}}_L(t,p) \right\rangle_t.
    \label{T(p)}
\end{gather}
With this transmission coefficient, we get the conventional Landauer formula for the average current:
\begin{align}
    \Big\langle\langle \hat{I}(x,t) \rangle \Big\rangle_{t} = \frac{e v}{h} \int dp\, \mathcal{T} (p)
    \Big[ f_L(p) - f_R(p) \Big].
    \label{I-Landauer}
\end{align}

Next, we derive the expression for the correlation function of current fluctuations in the left lead in the \textit{inelastic scattering} case: 
\begin{widetext}
\begin{align}
\label{currentCorrInelas}
&\langle \delta\hat{I}(x,t) \delta\hat{I}(x',t') \rangle = \left(\frac{\hbar e}{4\pi m}\right)^2 
\int dk\,dk'\,dp\,dp'\,dq\,dq' 
\e^{i(E_{k'} -E_k) t + i (E_{q'} -E_q) t'} 
\\
\nonumber
&\times 
\bigg[ 
f_L(p) [1-f_L(p')]
C_{LL}\left(k,k';p,p';x\right)
C_{LL}\left(q,q';p',p;x'\right)
 + 
f_R(p) [1-f_R(p')]
C_{RR}\left(k,k';p,p';x\right)
C_{RR}\left(q,q';p',p;x'\right)
\\
&\,\, + 
f_L(p) [1-f_R(p')]
C_{LR}\left(k,k';p,p';x\right)
C_{RL}\left(q,q';p',p;x'\right)
 + 
f_R(p) [1-f_L(p')]
C_{RL}\left(k,k';p,p';x\right)
C_{LR}\left(q,q';p',p;x'\right) \bigg].
\nonumber
\end{align}
The noise power in the inelastic case is define through the time-average of the correlation function (\ref{currentCorrInelas}) as follows:
\begin{gather}
\label{sPowerDef}
    S(\omega = 0) = \int dt' 
    \Big\langle \langle \delta\hat{I}(x,t + t') \delta\hat{I}(x',t) \rangle \Big\rangle_{t}. 
\end{gather}
Substituting Eq.~(\ref{currentCorrInelas}) into Eq.~(\ref{sPowerDef}) and using Eqs.~(\ref{CLL})-(\ref{CRL}), (\ref{ttp}) and (\ref{rtp}), we obtain a general expression for the current noise in the left lead in terms of inelastic scattering amplitudes:
\begin{align}
&S_L = S_{\text{therm}} + S_{\text{shot}}, 
\label{St+Ss}
\\ 
&S_{\text{therm}} = \frac{e^2 }{h} \int dk\,v\, 
\Big\langle \tilde{\texttt{t}}_L^*(t,k) \tilde{\texttt{t}}_L(t,k) \Big\rangle_t 
\bigg\{ f_L(k) [1-f_L(k)] + f_R(k) [1-f_R(k)] \bigg\},  
\label{Stherm}
\\  
&S_{\text{shot}} = \frac{e^2 }{h} \int dp\,dp'\,dk\,v\, 
\label{sInelGeneral2}
\\
&\times \bigg\{
f_L(p) [1-f_R(p')]\, \texttt{r}_L^*(k,p)\texttt{t}_L(k,p') \,
\Big\langle\tilde{\texttt{t}}_L ^*(t,p') 
\tilde{\texttt{r}}_L(t,p) \Big \rangle_t 
+ 
 f_R(p) [1-f_L(p')]\, \texttt{t}_L^*(k,p)
\texttt{r}_L(k,p')\,
\Big\langle \tilde{\texttt{r}}_L^*(t,p') \tilde{\texttt{t}}_L(t,p) \Big\rangle_t 
\notag
\\
 &- f_L(p) [1-f_L(p')]\, \texttt{r}_L^*(k,p)\texttt{r}_L(k,p') \,
\Big\langle\tilde{\texttt{t}}_L ^*(t,p') 
\tilde{\texttt{t}}_L(t,p) \Big \rangle_t 
-
f_R(p) [1-f_R(p')]\, \texttt{r}_L^*(k,p)
\texttt{r}_L(k,p')\,
\Big\langle \tilde{\texttt{t}}_L^*(t,p') \tilde{\texttt{t}}_L(t,p) \Big\rangle_t \bigg\}. 
\notag
\end{align}
\end{widetext}
Here, we split the total noise into the thermal part $S_\text{therm}$, which vanishes exactly when the temperatures of the reservoirs are sent to zero, and the remaining shot-noise part $S_\text{shot}$, which remains finite in this limit. 
When the scattering is elastic, i.e., when $\texttt{t}_j(k,p) = \delta(k-p) \texttt{t}(p)$ and the same holds for the reflection amplitudes, Eqs.~(\ref{St+Ss})-(\ref{sInelGeneral2}) reduce to Eq.~(\ref{sElGeneral}). In particular, when contacts are kept at $T=0$, Eq.~(\ref{sInelGeneral2}) becomes, for elastic scattering,  Eq.~(\ref{zero-T-elastic-shot}).

For the noise in the right lead, one performs a replacement $R\leftrightarrow L$, as usual. Equations (\ref{St+Ss})-(\ref{sInelGeneral2}) is the central result of this work. In what follows, it will be used to calculate the shot noise in a double-barrier structure in the presence of inelastic electron-phonon scattering.

\section{Transmission coefficient with dephasing}
\label{Sec:II}

In Sec.~\ref{Sec:II}, we have derived a general formula, Eq.~(\ref{T(p)}), for the transmission coefficient describing an inelastic scatterer. 
In this section, we will obtain an expression for the transmission coefficient for a one-dimensional double-barrier structure in the presence of a time-dependent potential. We will then use this result to compute the transmission coefficient for the case of inelastic scattering due to electron-phonon interaction. This will allow us to calculate the conductance of the structure and to analyze the role of phonon-induced dephasing in resonant tunneling. As a warm-up, we analyze in Appendix~\ref{Appendix1} the case of a time-dependent single barrier modelled by a delta-function potential. Next, we generalize that
consideration to the spatially extended double-barrier setup, and introduce two microscopic models of phonon-induced dephasing.

\subsection{Transmission coefficient for a double barrier structure with time-dependent potential}

We have considered a simple example of transmission through a time-dependent delta-barrier in Appendix~\ref{Appendix1} and derived the inelastic transmission amplitude $\texttt{t}(p,k)$ for this ``toy model''. In this section,  we obtain the transmission coefficient $\mathcal{T}(p)$ of a 1D quantum dot with tunnel contacts in the presence of a random time-dependent potential $V(x,t)$, and then perform averaging over realizations of the phonon-induced potential.

We assume that the random potential is applied only inside the quantum dot formed by the two barriers. We further assume that the magnitude of the potential is smaller than the electron's kinetic energy. Next, the potential is considered to be smoothly varying both in space and in time.
This will allow us to neglect the electron backscattering induced by the potential 
($r_c k \gg 1$, where $k$ is the wave vector of  an electron and $r_c$ is a characteristic spatial scale of the potential) and transitions between the levels of size quantization inside the dot:
$$\partial V(x,t) / \partial t \ll V(x,t) / \tau_f, $$  
where $\tau_f=2L/v$  is the time of flight back and forth between the barriers separated by distance $L$. 

Under these assumptions and upon linearization of the electron's dispersion, $E_k \approx  \hbar v k$, the influence of the random potential on the electron wave function reduces to the appearance of a random phase factor:
\begin{align}
\label{psiDep}
\psi(x,t) &= \exp\left(ikx - i \frac{E}{\hbar} t\right)
\\
&\times
\exp\left[ -\frac{i}{\hbar}\int^t\! d\tau\  V(x - v t + v \tau, \tau)\right], 
\nonumber
\\[0.3cm] 
\label{dephCond}
\text{for}\quad&\Delta\, \tau_c \gg \hbar, \quad k\, r_c \gg 1, \quad |V(x,t)| \ll \hbar v k,
\end{align}
where 
$$\Delta = 2\pi \hbar / \tau_f$$ 
is the interlevel energy spacing inside the ``quantum dot'' formed by the barriers.
We can introduce here the already mentioned before  random addition to electron wave function phase:
$$\varphi_f = \frac{1}{\hbar}\int^t\! d\tau\  V(x - v t + v \tau, \tau).$$
With this form for the wave function, we obtain a general expression for transmission amplitude for a double barrier structure in the presence of a given realization of weak random potential (\ref{dephCond}). The two pointlike barriers labelled by $i=1,2$ are located at $x=0$ and $x=L$ and are characterized by their individual transmission and reflection amplitudes,  $\texttt{t}_i$ and $\texttt{r}_i$, respectively (for each of the barriers, its right and left reflection amplitudes are equal). As shown in Appendix \ref{Appendix2}, the inelastic transmission amplitude $\texttt{t}_R$ for electrons transmitted to the right lead is given by
\begin{widetext}
	\begin{gather} 
	\label{tPKGen}
	\texttt{t}_R(p,k) = \int_{-\infty}^\infty dy\, \e^{i(k-p)y}  
	\texttt{t}_1\texttt{t}_2 
	\sum_{n=0}^\infty \left( \texttt{r}_1 \texttt{r}_2 \e^{2i kL} \right)^{n}  
	\exp \left[
    -\frac{i}{\hbar} \int_{-\tau_f n}^0 d\tau\, V(x_\text{in}(\tau), \tau - y/v)
    \right], 
    \quad \tau_f = 2L/v,
    \end{gather}
where $x_\text{in}(t)$ is the trajectory of a particle between the barriers, which consists of ballistic segments of length $L$. Using Eq.~(\ref{tPKGen}), the transmission coefficient (\ref{T(p)}) takes the following form:     \begin{gather}
	\mathcal{T}(k) = \langle \texttt{t}^*_R(t,k) \texttt{t}_R(t,k) \rangle_t = |\texttt{t}_1|^2 |\texttt{t}_2|^2 \sum_{n_1,n_2} \left(\texttt{r}_1\texttt{r}_2 \e^{2ikL}\right)^{n_1} \left(\texttt{r}_1^*\texttt{r}_2^* \e^{-2ikL}\right)^{n_2} 
	\left\langle
	\exp\left[-\frac{i}{\hbar}\int^{-\tau_f n_2}_{-\tau_f n_1} d\tau V(x_{\text{in}}(\tau),\tau + t)
	\right] \right\rangle_t, 
	\label{tGenV}
	\end{gather}
\end{widetext}
The transmission coefficient (\ref{tGenV}) in the absence of dephasing potential reduces to the well-known expression 
$$\mathcal{T}(k) = \frac{|\texttt{t}_1\texttt{t}_2|^2}{|1-\texttt{r}_1\texttt{r}_2\e^{2ikL}|^2}.$$

Now the task is to perform the averaging of the exponential factor in Eq.~(\ref{tGenV}) over the random field. For the Gaussian distribution of $V(x,t)$, the averaging can be done exactly. Furthermore, assuming that the characteristic correlation time $\tau_c$ for the variation of $V(x,t)$ is longer than the flight time $\tau_f$, one can first average $V(x_{\text{in}}(\tau),\tau + t)$ over the position between the barriers, yielding an effective time-dependent potential $U(\tau+t)$. %
Then the averaging over fluctuations of this potential gives rise to the dephasing factor, 
\begin{align}
\label{expCalc}
&f(t) \equiv \av{\e^{i \int_{0}^{t}d\tau U(\tau)}} \\ 
&\ =\exp \left\{\frac{\left[\int_{0}^{t}d\tau K(\tau)\right]^2}{2K(0)} - \frac{1}{2}\int_0^t\int_0^t d\tau d\tau' K(\tau-\tau')\right\}
\notag
\\
\label{eAvLim}
&\ = 
\begin{cases}
1, \quad & t \ll \tau_c, \\
\e^{ - \int_{0}^{t} \int_{0}^{t} d\tau d\tau' K(\tau-\tau')/2}, \quad & t \gg \tau_c.
\end{cases}
\end{align}
expressed through the correlation function 
\begin{align} 
\label{kU}
&K(\tau-\tau') = \av{U(\tau)U(\tau')}_\text{U},
\end{align}
where averaring is performed over the random potential fluctuations. In the case of the phonon potential, this implies averaging over the random phases of phonon modes (in the classical approach), or with the phonon density matrix (quantum).

Dephasing processes can affect the transmission coefficient only when the correlation time of the potential fluctuations is smaller than time $\tau_{\text{in}}$ that the particle spends between the barriers, i.e. time flight multiplied by resonator quality factor.
Therefore, we further assume the following time hierarchy: 
$$\tau_f \ll \tau_c \ll \tau_{\text{in}},$$ and use, in what follows, the second line of Eq.~(\ref{eAvLim}) for $f(t)$. 

The applicability of Eq.~(\ref{tPKGen}) in a parabolic spectrum is limited by the use of a quasiclassical wave function corresponding to the linearized dispersion and classical external potential $V(x,t)$. This assumes that the change of particle's velocity $v$ is small in the process of the backscattering off the barriers in the presence of a fluctuating field, leading to the ``heating'' or ``cooling'' of the particle, in addition to dephasing. 
The corresponding condition can be expressed through the quality factor $Q$ of the resonator, as follows:
$$Q \ll \frac{v \tau_c}{\text{min}[r_c,L]} \left(\frac{E}{V}\right)^2,$$ 
where $V$ is the characteristic magnitude of the fluctuating potential.
Indeed if the resonator's quality factor is very high and the particle is influenced by the random potential for a very long time, its speed (and, hence, energy) may change. When considering the case of electron-phonon interaction, a rather small rate of quantum spontaneous phonon emission is also required, in order to justify the quasiclassical approach.

\subsection{Transmission for the diffusion type of dephasing}

Equation (\ref{expCalc}) can be simplified in the case of diffusion type of phase dynamics, when $K \tau_c^2 / \hbar^2 \ll 1$, where $K = K(0)$ is the square of a characteristic magnitude of the effective random potential, see Eq.~(\ref{kU}). In this regime, the random-sign changes, $\delta\varphi_f$, of the phase of the electron wave function, caused by the potential during time $\tau_c$, are small: 
$\delta \varphi_f \ll 1$. 
The total change of the electronic phase $\varphi_f$, i.e., the sum of these random elementary changes, will then show a phase diffusion at long times:
$$\langle[(\varphi_f(t)-\varphi_f(0)]^2\rangle\propto t.$$ 

For $t \gg \tau_c$ in the diffusion regime of dephasing, we can rewrite the dephasing factor, Eq.~(\ref{eAvLim}), in the conventional exponential form:
\begin{gather}
\label{expDiff}
f(t) \approx \e^{-t/\tau_\varphi}.
\end{gather}
Here we introduced the dephasing time $\tau_\varphi$ according to
\begin{gather}
\frac{1}{\tau_\varphi} =  
\int_0^\infty \frac{d\tau}{\hbar^2} K (\tau).
\label{tauvarphi}
\end{gather}
The inverse dephasing time is the phase diffusion coefficient in this case.
It may turn out that the time integral of the correlator (\ref{kU}) that enters Eq.~(\ref{tauvarphi}) vanishes. We will see below that this situation is realized for the phonon-induced potential. In this case, as we will show below, one needs to calculate exactly the double time integral entering  (\ref{expCalc}). 

Now, let us calculate the transmission coefficient (\ref{tGenV}) for the diffusion type of phase dynamics (\ref{expDiff}).
In this case, the transmission coefficient can be calculated exactly:
\begin{widetext}
	\begin{gather}
	\label{tDyn}
	\mathcal{T} (k) 
	=
	\frac{|t_1|^2|t_2|^2 \left[1-|r_1|^2|r_2|^2\exp(-2 \tau_f/ \tau_\varphi)\right]}{(1-|r_1|^2|r_2|^2)\left[1-r_1 r_2 \exp(2ikL-\tau_f / \tau_\varphi)\right] \left[1-r_1^* r_2^* \exp(-2ikL-\tau_f / \tau_\varphi)\right] }, 
	\end{gather} 
\end{widetext}
Equation (\ref{tDyn}) can be expanded near one resonant level, where it acquires the following, valid for an arbitrary dephasing rate:
\begin{gather}
\label{tExpandGen}
\mathcal{T} (\delta E) = \frac{\Gamma_1\Gamma_2}{\Gamma_\Sigma}\frac{ \frac{\hbar}{\tau_f}\sinh(\tau_f/\tau_\varphi) + \Gamma_\Sigma\, e^{-\tau_f/\tau_\varphi}}{\left[  \frac{2\hbar}{\tau_f}\sinh\left(\frac{\tau_f}{2\tau_\varphi} \right) +\Gamma_\Sigma\, \e^{-\tau_f/2 \tau_\varphi} \right]^2 + \delta E^2}.
\end{gather}  
Here, $\Gamma_1 = |t_1|^2 \hbar/\tau_f$ and $\Gamma_2 = |t_2|^2 \hbar/\tau_f$ are the transmission coefficients of individual barriers, $\Gamma_\Sigma = (\Gamma_1 + \Gamma_2)/2$, and $\delta E$ is the deviation of the electron's energy from the resonance (we assumed a small transparency for both barriers, $|t_i| \ll 1$).

Transmission across the double barrier structure can be considered in three regimes depending on the dephasing rate. 
If dephasing rate is small, $\tau_{\text{in}} \sim \tau_\varphi$, the ``resonant coherent tunneling'' is realized, and we can transform Eq.~(\ref{tExpandGen}) into the conventional Breit-Wigner form:
\begin{gather}
\label{TLee}
\mathcal{T} (\delta E) = \frac{\Gamma_1 \Gamma_2}{\Gamma_\Sigma} \frac{\Gamma_\Sigma + \hbar/ \tau_\varphi}{(\Gamma_\Sigma + \hbar/ \tau_\varphi)^2 + \delta E^2}.
\end{gather} 
Such a dephasing-broadened Breit-Wigner resonance was first obtained for this model by Stone and Lee \cite{Lee_PRL}. It is seen that the resulting quasi-level's width consists of two part: 
$$\tilde{\Gamma} = \Gamma_\Sigma + \hbar / \tau_\varphi,$$ 
governed by the transparency of barriers and the dephasing rate.

In the case of strong dephasing $\hbar / \tau_\varphi\sim \Delta$, the regime of ``classical tunneling'' sets in. 
In this case, particle transmit across the double-barrier structure like a classical object, and the total transmission coefficient is determined by the classical transmission probabilities of individual barriers. It is worth noting that in this regime Eq.~(\ref{tDyn}) is formally not applicable, since transitions between the size-quantization levels start to play an important role in the quantum-mechanical description. However, Eq.~(\ref{tDyn}) still yields a correct classical result for $\mathcal{T}(k)$:
\begin{gather}
\label{tClassic}
\mathcal{T} = \frac{|t_1|^2|t_2|^2}{1-|r_1|^2|r_2|^2}
=\frac{\mathcal{T}_1\mathcal{T}_2 }{\mathcal{T}_1+\mathcal{T}_2-\mathcal{T}_1\mathcal{T}_2}
\end{gather}
The result is obtained from Eq.~(\ref{tDyn}) and is valid for an arbitrary transparency of the barriers. For strong barriers, one can use directly Eq.~(\ref{tExpandGen}), which, in the limit of strong dephasing,
yields
\begin{gather}
\mathcal{T}
\approx \frac{\Gamma_1 \Gamma_2}{\Gamma_1 + \Gamma_2} \frac{2 \pi}{ \Delta}.
\label{T}
\end{gather}  
For intermediate values of dephasing rate, $1/\tau_f > 1 / \tau_\varphi > 1/\tau_{\text{in}}$, the regime of ``coherent sequential tunneling'' is realized, where the transmission coefficient is described by Eq.~(\ref{tExpandGen}). 

\subsection{Transmission for non-diffusive dephasing}

In some cases, the diffusion approximation for the phase dynamics may fail. In particular, for the case of electron-phonon interaction considered in detail in Sec.~\ref{Sec:IV} below, logarithmic phase dynamics, 
$$\langle [\varphi_f(t)-\varphi_f(0)]^2 \rangle = \gamma \ln (t / \tau_c),$$
occurs for $t > \tau_c$ instead of phase diffusion.
This gives rise to a power-law time dependence of the dephasing factor in Eq.~(\ref{expCalc}):
\begin{gather}
    f(t) \approx \e^{-\gamma \ln(t/\tau_c)} = \left( \frac{\tau_c}{t} \right)^\gamma.
    \label{expLogDef}
\end{gather}
We will refer to this type of dephasing as  ``logarithmic dephasing''.

Now, we calculate the transmission coefficient for the logarithmic type of phase dynamics (\ref{expLogDef}). The calculation is a somewhat less straightforward.
First, Eq.~(\ref{tGenV}) with the power-law dephasing factor (\ref{expLogDef}) yields
	\begin{align}
		\mathcal{T} (k) &= |t_1|^2|t_2|^2 \sum_{n_1, n_2} \left(r_1 r_2 \e^{2ikL}\right)^{n_1} \left(r_1^* r_2^* \e^{-2ikL}\right)^{n_2} 
		\notag
		\\
	&\times
	\left(\frac{1}{|n_1-n_2| \tau_f/\tau_c + 1}\right)^{\gamma}. 
	\label{TnonDifSum}
\end{align}
Near the resonance, the summation over $n_1$ and $n_2$ in Eq.~(\ref{TnonDifSum}) can be performed, leading to
\begin{align}
	\label{Tln}
	\mathcal{T} (\delta E) = \frac{\Gamma_1 \Gamma_2 \tau_c}{2\hbar} \left[ \e^\lambda \mathcal{E}_\gamma(\lambda) + \e^{\lambda^*} \mathcal{E}_\gamma(\lambda^*) \right], 
\end{align}	
where we introduced
\begin{align} 
	\lambda = \frac{\Gamma_\Sigma+i\delta E}{\hbar/\tau_c}
\end{align}
and
\begin{align}
	\mathcal{E}_\gamma(\lambda) = \int_1^\infty dx\, x^{-\gamma} \e^{-\lambda x}
	\label{ExpInt}
\end{align}
is the exponential integral function.
Specifically, in order to calculate the sum in Eq.~(\ref{TnonDifSum}) we used the following integral representation of a power-law function:
\begin{gather}
\label{powerInt}
\frac{1}{z^\gamma} = \frac{1}{\Gamma(\gamma)} \int_0^\infty dx\, x^{\gamma - 1} \e^{-z x},
\end{gather}
with $\Gamma(\gamma)$ the Gamma function. 
This amounts to introducing an effective 
exponential dephasing factor (\ref{expDiff}) with $\tau_\varphi$ depending on $x$, followed by averaging over $x$.
By using this, we calculated the sum in Eq.~(\ref{TnonDifSum}) for a given value of $x$, exactly as it was done for the exponential dephasing factor in  Eq.~(\ref{tDyn}). 
The last integral over $x$ in Eq.~(\ref{powerInt}) can be calculated exactly for one size quantization level. We expanded the integrand near the resonant level (\ref{TLee}), multiply it with $x^{-\gamma}$, and obtain Eq.~(\ref{Tln}) after the integration over $x$. 
The integral can be converted to the integral (\ref{ExpInt}) which starts from unity instead of zero. 

\begin{figure}
	\centering
	\includegraphics[width=\linewidth]{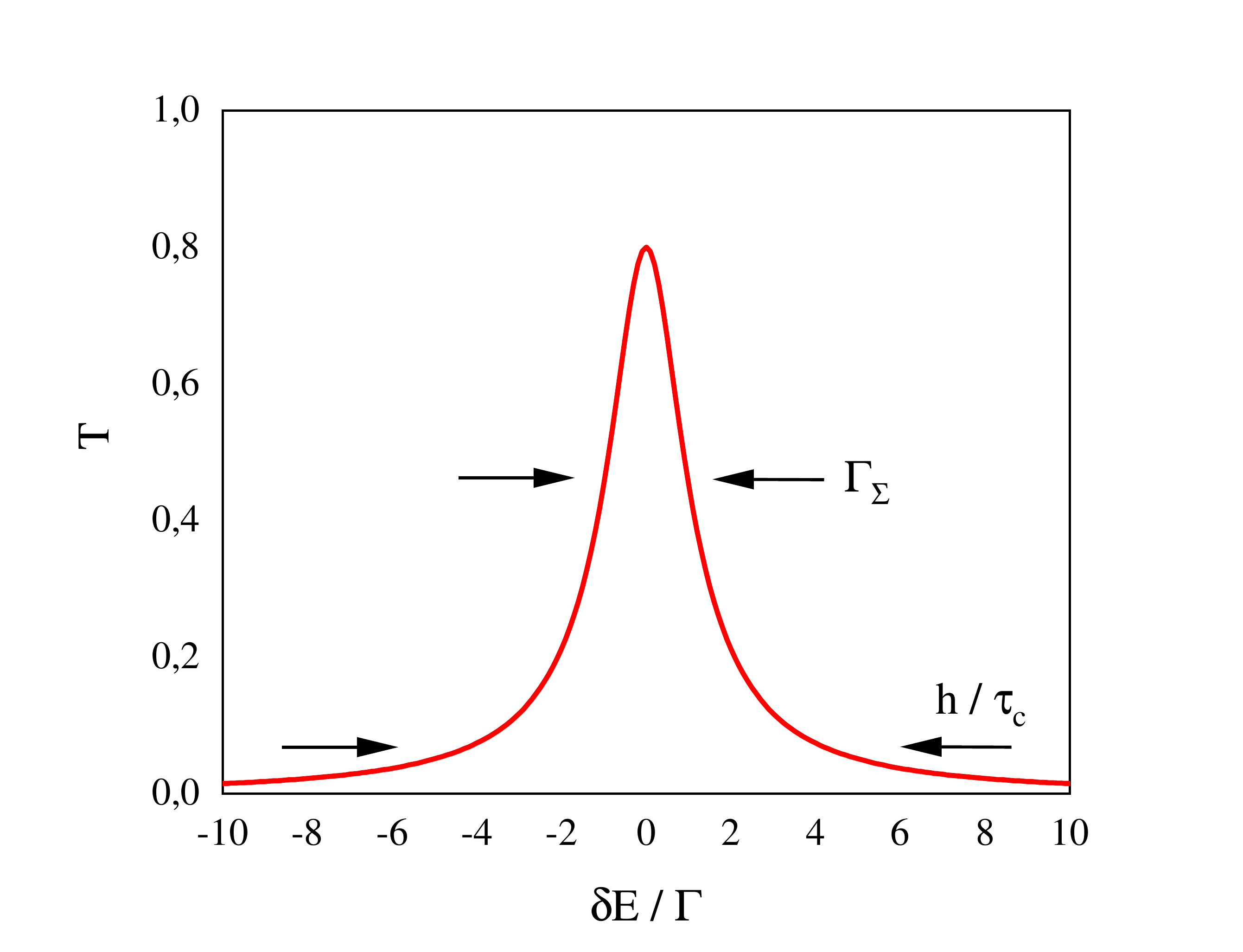}
	\caption{Transmission coefficient (\ref{Tln}) for the logarithmic dephasing has two energy scales: one is determined by the transparency of barriers through $\Gamma_\Sigma$ and the other by the correlation time $\tau_c$.}
	\label{figT}
\end{figure}

The transmission coefficient (\ref{Tln}) for logarithmic dephasing is not described by a standard Lorentzian Breit-Wigner energy dependence. Instead, it is caracterized by two energy scales, $\Gamma_\Sigma$ and $\hbar/\tau_c$, see Fig.~\ref{figT}. Assuming $\gamma \ll 1$, we can simplify Eq.~(\ref{Tln}):
\begin{gather}
\label{transNonDiff}
\mathcal{T} (\delta E) \approx \frac{\Gamma_1 \Gamma_2 \tau_c}{2\hbar} \left[ \left(\frac{\hbar/\tau_c}{\Gamma_\Sigma+i\delta E}\right)^{1-\gamma}\!\!+\! \left(\frac{\hbar/\tau_c}{\Gamma_\Sigma-i\delta E}\right)^{1-\gamma} \right].
\end{gather} 
For energies $\delta E \gg \Gamma_\Sigma$, the transmission coefficient gradually decreases on scale $\hbar / \tau_c$. 

When $K \tau_c^2 / \hbar^2 \gg 1$, yet another regime of wave function phase growing realized, which we term ``ballistic regime''. This means that the random change of the wave-function phase is large, $\delta \varphi_f \gg 1$, on the potential correlation time scale. 
In the ballistic case, the dephasing exponent in the dephasing factor is given
by
\begin{gather}
\nonumber
	\ln\,f(t) \approx - \frac{t^2 K (0)}{2 \hbar^2},
	\label{lnFBal}
\end{gather}
which leads to the characteristic phase  decay rate $$1/\tau_\varphi \approx \sqrt{K(0)}/\hbar.$$

\subsection{Conductance}
The linear conductance across system is calculated by means of the Landauer formula that follows from Eq.~(\ref{I-Landauer}):
\begin{gather}
G = - \frac{e^2}{h} \int dE\, \frac{\partial f}{\partial E} 
\langle \mathcal{T}(E) \rangle,
\end{gather}   
where $f$ is the Fermi-Dirac distribution function. 
At low temperatures $T \ll \tilde{\Gamma} \ll \Delta$, one finds:
\begin{gather} \label{gTlow}
G = \frac{e^2}{h} \sum_n \frac{\Gamma_1 \Gamma_2}{\Gamma_\Sigma} \frac{\tilde{\Gamma}}{\tilde{\Gamma}^2 + (E_n - \mu)^2},
\end{gather} 
where $\mu$ is the Fermi level in the leads and $E_n$ is the energy of the $n$th quasi-level. 
Here, one needs to make a remark that, for logarithmic dephasing, the result may in general depend on the ratio of $T$ and $\hbar/\tau_c$.
If $T \ll \hbar/\tau_c$, one has to use in Eq.~(\ref{gTlow}) the transmission coefficients in the form of Eq.~(\ref{transNonDiff}). However, for the relevant case of phonon-induced dephasing, as we show later in Sec.~\ref{Sec:IV}, the correlation time $\tau_c$ is determined by the temperature, $\hbar / \tau_c = T$. As a consequence, for $T \ll \Gamma_\Sigma$ the transmission coefficient (\ref{Tln}) results in Eq.~(\ref{gTlow}) with replacing $\tilde{\Gamma} \rightarrow \Gamma_\Sigma$.

If temperature exceeds the level width, $\tilde{\Gamma} \ll T \ll \Delta$, we get:
\begin{gather} \label{gThigh}
G = \frac{e^2}{h} \frac{\Gamma_1 \Gamma_2}{\Gamma_1 + \Gamma_2} \sum_n \frac{\pi}{2 T \cosh^2\left(\frac{E_n - \mu}{2T}\right)}.  
\end{gather}
Note that the conductance is independent of the dephasing rate, and hence on the mechanism of dephasing, already in this regime. For temperatures that are higher than the level spacing, $\Delta \ll T$, the conductance reads:
\begin{gather}
\label{gDiff}
G =  \frac{e^2}{h} \frac{\Gamma_1 \Gamma_2}{\Gamma_1 + \Gamma_2} \frac{2 \pi}{ \Delta}. 
\end{gather}  
This result can also be obtained by means of the classical consideration that corresponds to Eq.~(\ref{T}) for the ``classical'' transmission coefficient.

\section{Shot noise}
In this section, we analyze shot noise in the double barrier structure. As we mentioned before, the Fano factor is defined by shot noise at zero temperature. Therefore, in this section we will assume that the lead are kept at $T = 0$. Also to simplify equations we assume that applied voltage is $2V$. 
We start from the fully coherent case:
\begin{align}
    &F = \frac{\int_{\mu-eV}^{\mu+eV} dE\, \mathcal{T}(E)[1-\mathcal{T}(E)]}{\int_{\mu-eV}^{\mu+eV} dE\, \mathcal{T}(E)} 
     \label{fanoSimpleV} \\
    &= \frac{\Gamma_1^2 + \Gamma_2^2}{(\Gamma_1 + \Gamma_2)^2} - \frac{2 \Gamma_1 \Gamma_2}{(\Gamma_1 + \Gamma_2)^2} 
    \frac{eV \Gamma_\Sigma}{\Gamma_\Sigma^2+(eV)^2} \frac{1}{\arctan(eV/\Gamma_\Sigma)}.
   \notag
\end{align}
Here and in what follows, we have assumed that the Fermi level is exactly at one of the size quantization levels. 
This equation is valid irrespective of the relation between $eV$ and $\Delta$. 

For the inelastic case, in the presence of random potential, it is useful to write the numerator of Eq.~(\ref{fanoSimpleV}), i.e., shot noise (\ref{sInelGeneral2}), using the sum representation. As shown in Appendix \ref{Appendix2}, reflection amplitudes for the double barrier structure 
can be represented as
\begin{gather}
\label{rtRel}
    \texttt{r}_{R,L}(k,p) = \frac{\texttt{r}_2 \e^{2ikL} -\texttt{r}_1^* }{\texttt{t}_1^* \texttt{t}_2} \texttt{t}_{L,R}(k,p).
\end{gather}
Using this, we have for $T=0$:
\begin{align}
\notag
    S_L &= \frac{e^2}{h} \int_{(\mu-eV)/v}^{(\mu+eV)/v} dk\,v\, 
    |\texttt{r}_1 \e^{ikL} - \texttt{r}_2^* \e^{-ikL}|^2 \\ 
   &\times \sum_{\{n_i\}} 
    \left(\texttt{r}_1\texttt{r}_2 \e^{2ikL}\right)^{n_1 + n_3} \left(\texttt{r}_1^*\texttt{r}_2^* \e^{-2ikL}\right)^{n_2 + n_4}
    \notag
    \\
   &\times \left\langle\exp\left(-\frac{i}{\hbar}\int^{\tau_f (n_1-n_2)}_{\tau_f (n_4-n_3)} d\tau U(\tau) \right)
    \right\rangle_t.
    \label{sSumV}
\end{align}
For $U(\tau)=0$, we immediately arrive at Eq.~(\ref{fanoSimpleV}).

Let us now analyze the shot noise for the diffusion type of dephasing:
\begin{gather}
\notag
    \left\langle\exp\left(-\frac{i}{\hbar}\int^{\tau_f (n_1-n_2)}_{\tau_f (n_4-n_3)} d\tau U(\tau) \right)
    \right\rangle_t \\
    =\exp\left(-\frac{\tau_f}{\tau_\varphi} |n_1-n_2+n_3-n_4|\right).
\end{gather}
We can represent this exponential factor in as follows:
\begin{gather}
    \exp(-|n| \tau_f/\tau_\varphi) = \int \frac{dy}{\pi}\frac{\tau_f/\tau_\varphi}{(\tau_f/\tau_\varphi)^2 + y^2} 
    \e^{i n y}.
\end{gather}
Next, we use $\Gamma_\varphi = \hbar / \tau_\varphi$ to simplify equations. 
With this representation we can use the expression for shot noise in coherent case [i.e., Eq.~(\ref{sSumV}) with $U(\tau)=0$], replace there $k \rightarrow k + y$ and perform the integration over $\delta=y \hbar/\tau_f$:
\begin{align}
\notag
    S_L& = \frac{e^2}{h} \int_{\mu-eV}^{\mu+eV} dE \int_{-\infty}^\infty \frac{d\delta}{\pi}
    \frac{\Gamma_\varphi}{\Gamma_\varphi^2 + \delta^2}
    \\
    &\times 
    \mathcal{T}(E+\delta) [1-\mathcal{T}(E+\delta)]. 
    \label{SL-diff}
    \end{align}
This results in the  following Fano factor
for diffusive dephasing:
    \begin{align}
       F_{\text{diff}}& = \frac{\Gamma_1^2 + \Gamma_2^2}{(\Gamma_1 + \Gamma_2)^2}
     \label{FanoDiffExact}  
     \\ 
       &- \frac{2 \Gamma_1 \Gamma_2}{(\Gamma_1 + \Gamma_2)^2} 
    \frac{eV \Gamma_\Sigma}{(\Gamma_\Sigma+\Gamma_\varphi)^2+(eV)^2} \frac{1}{\arctan\left(\frac{eV}{\Gamma_\Sigma+\Gamma_\varphi}\right)}.
    \notag
\end{align}
Compared with the result 
for the coherent case, Eq.~(\ref{fanoSimpleV}), the diffusive dephasing leads to the replacement $\Gamma_\Sigma\to \Gamma_\Sigma+\Gamma_\varphi$ everywhere except for the numerator of the second term.

\begin{figure}[t]
	\centering
	\includegraphics[width=\linewidth]{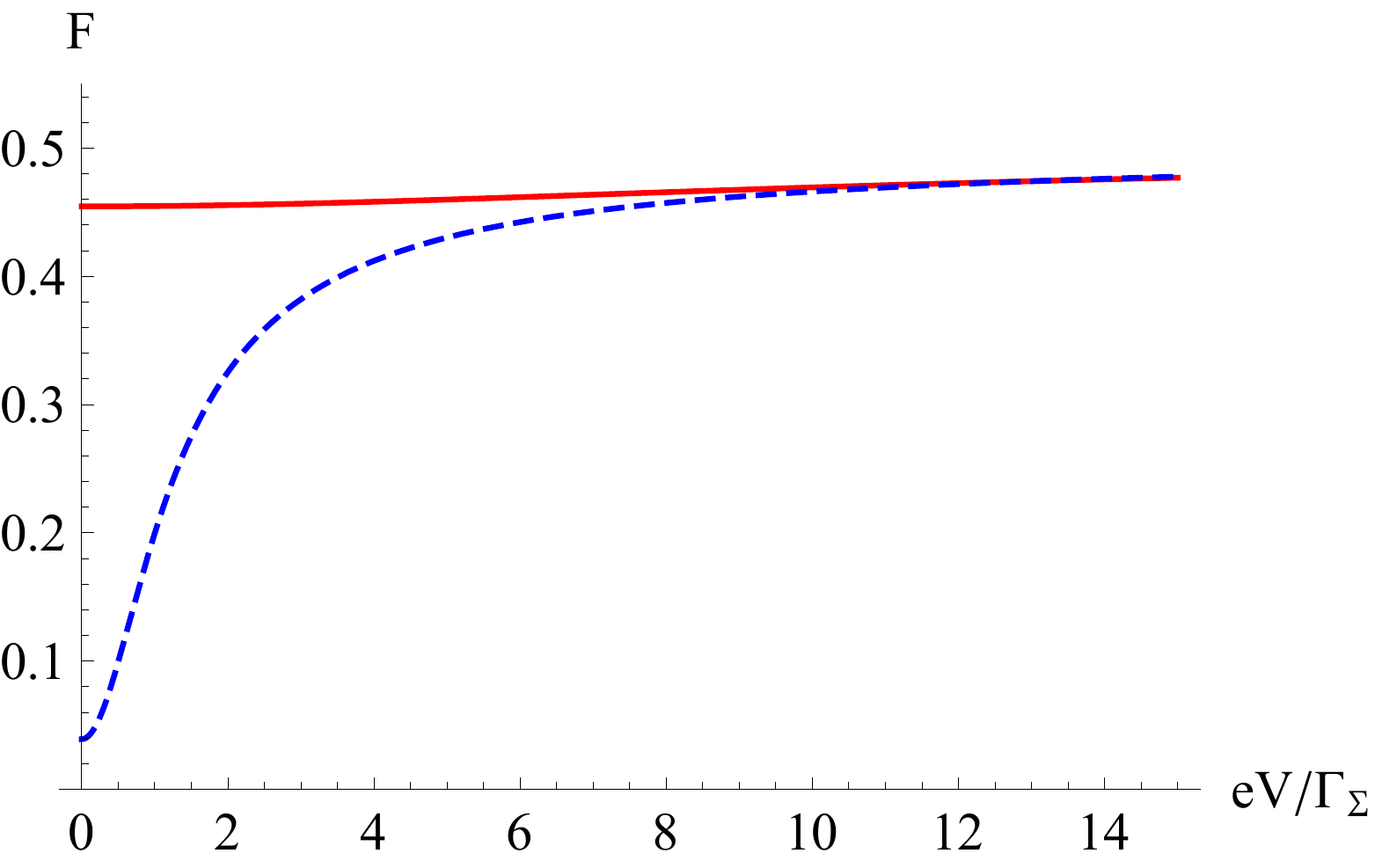}
	\caption{The Fano factor for different types of dephasing as a function of applied voltage. Red solid line is for the diffusion type of dephasing, Eq.~(\ref{FanoDiffExact}); blue dashed line is for the logarithmic type, Eq.~(\ref{FanoLogExact}). Parameters for the plot: $\Gamma_\varphi/\Gamma_\Sigma = 10$ and $\gamma = 0.1$.}
	\label{fanoPlot}
\end{figure}

Now we calculate the shot noise the case of the logarithmic dephasing. For this we use the same trick as for the calculation of the transmission coefficient:
\begin{align}
\notag
    &\left\langle\exp\left(-\frac{i}{\hbar}\int^{\tau_f (n_1-n_2)}_{\tau_f (n_4-n_3)} d\tau U(\tau) \right)
    \right\rangle_t \\ \notag
    &=\left(\frac{1}{ |n_1-n_2+n_3-n_4|\tau_f / \tau_c +1}\right)^\gamma  = \int_0^\infty\!\! d y \frac{y^{\gamma-1}\e^{-y}}{\Gamma(\gamma)} \\
    &\times 
    \int \frac{dx}{\pi}
    \frac{y \tau_f/\tau_c}{(y\tau_f/\tau_c)^2 + x^2} 
    \e^{i (n_1-n_2+n_3-n_4) x}.
\end{align}
For logarithmic dephasing, we can introduce the dephasing strength, similarly to the diffusion type: 
$$\Gamma_\varphi = \hbar / \tau_c.$$ 
Then, for the calculation of the Fano factor, we can perform the same procedure as for diffusion type of phase dynamics:
\begin{gather}
\label{FanoLogExact}
    F_{\text{log}} = \frac{\Gamma_1^2 + \Gamma_2^2}{(\Gamma_1 + \Gamma_2)^2} 
    - \frac{2 \Gamma_1 \Gamma_2}{(\Gamma_1 + \Gamma_2)^2} 
    \frac{G}{H}, \\
    G = \int \frac{d y}{\Gamma(\gamma)} y^{\gamma-1}\e^{-y} \frac{e V \Gamma_\Sigma}{(eV)^2 + (\Gamma_\Sigma + y \Gamma_\varphi)^2}, \\
    H = \int \frac{d y}{\Gamma(\gamma)} y^{\gamma-1}\e^{-y} \arctan \left(
    \frac{eV}{\Gamma_\Sigma + y \Gamma_\varphi}
    \right).
\end{gather}
Using these equations, we can distinguish three different regimes:
\begin{gather}
    \Gamma_\Sigma \ll \Gamma_\varphi \ll eV: \notag \\
    F_{\text{log}} = F_{\text{diff}} = \frac{\Gamma_1^2 + \Gamma_2^2}{(\Gamma_1 + \Gamma_2)^2} 
    - \frac{2 \Gamma_1 \Gamma_2}{(\Gamma_1 + \Gamma_2)^2} \frac{2}{\pi} 
    \left( \frac{\Gamma_\Sigma}{eV} \right), \\
        \Gamma_\Sigma \ll eV \ll \Gamma_\varphi: \notag \\
    F_{\text{log}} = \frac{\Gamma_1^2 + \Gamma_2^2}{(\Gamma_1 + \Gamma_2)^2} 
    - \frac{2 \Gamma_1 \Gamma_2}{(\Gamma_1 + \Gamma_2)^2} \frac{2}{\pi} 
    \left( \frac{\Gamma_\Sigma}{\Gamma_\varphi} \right)^\gamma \left( \frac{\Gamma_\Sigma}{eV} \right)^{1-\gamma}, 
    \label{F-log}
    \\
    F_{\text{diff}} = \frac{\Gamma_1^2 + \Gamma_2^2}{(\Gamma_1 + \Gamma_2)^2} 
    - \frac{2 \Gamma_1 \Gamma_2}{(\Gamma_1 + \Gamma_2)^2}  
     \frac{\Gamma_\Sigma}{\Gamma_\varphi}, \\
     eV \ll \Gamma_\Sigma \ll \Gamma_\varphi: \notag \\
     F_{\text{log}} = \frac{\Gamma_1^2 + \Gamma_2^2}{(\Gamma_1 + \Gamma_2)^2} 
    - \frac{2 \Gamma_1 \Gamma_2}{(\Gamma_1 + \Gamma_2)^2}, \\
    F_{\text{diff}} = \frac{\Gamma_1^2 + \Gamma_2^2}{(\Gamma_1 + \Gamma_2)^2} 
    - \frac{2 \Gamma_1 \Gamma_2}{(\Gamma_1 + \Gamma_2)^2}  
     \frac{\Gamma_\Sigma}{\Gamma_\varphi}.
\end{gather}
The main difference between the logarithmic and diffusion types of dephasing is now clearly seen. For the diffusion type of dephasing, the Fano factor is insensitive to the applied voltage when it is lower then the dephasing-induced contribution to the level width $\Gamma_\varphi$. For larger voltages, the Fano factor is given by the well-known result, with a small correction $\propto 1/ eV$. 
The Fano factor in the case of logarithmic dephasing has three different regimes depending on applied voltage. If the voltage is lower than the dephasing width $\Gamma_\varphi$ but higher than the elastic width $\Gamma_\Sigma$, the Fano factor depends on the voltage, in contrast to $F_\text{diff}$. This difference can be seen in Fig.~(\ref{fanoPlot}). In both cases, when the voltage exceeds the elastic width $\Gamma_\Sigma$, the Fano factor approaches the universal value (\ref{FanoSimple}) that can be obtained in the classical consideration. 

\section{Phonon induced dephasing}
\label{Sec:IV}

In this section, we will study the influence of acoustic phonons on the transmission and current fluctuations. We assume sufficiently low temperatures, when both backscattering and electron transitions between levels of quantization in the double-barrier structure due to their interaction with phonons can be neglected 
\footnote{For typical experimental systems, the velocity of longitudinal phonons is $s = 5 \cdot 10^5$~cm/s, the electron Fermi velocity $v = 10^7$~cm/s, and effective electron mass $m^* = 0.1\, m_0$, where $m_0$ is free electron mass. The absence of backscattering assumes
	$ k r_c \approx m^* v s/T \gg 1$, which is satisfied for $T \ll 5~\text{K}$. Absence of the transitions between the resonant levels requires 
	$ \tau_c \gg \tau_f, \, \text{or} \, L \ll v \tau_c(T = 1~\text{K}) = 1~ \mu \text{m}.$}.
Cases of one-dimensional (1D), two-dimensional (2D), and three-dimensional (3D) phonons will be considered. In experiment, all three possibilities can be realized.

Interaction with acoustic phonons is governed by the Hamiltonian:  
\begin{gather}
\hat{V}_\text{e-ph} = g \, \varphi({\bf{r}},t), \\ 
\varphi({\bf{r}},t) = \sum_{\bf{k}} \sqrt{\frac{\omega_k}{2 V}} \left( i b_{\bf{k}} \e^{i {\bf{kr}} - i \omega_k t} - i b_{\bf{k}}^* \e^{- i {\bf{kr}} + i \omega_k t} \right),
\end{gather}
where in the last equation only longitudinal phonons are included, $s$ is the speed of longitudinal phonons, and $\omega_k = s k$. Note that $g$ has different dimensions in the three cases. The average of phonon field amplitudes is given by 
$$\langle b_{\bf{k}} b_{\bf{p}}^* \rangle = \langle b_{\bf{p}}^* b_{\bf{k}} \rangle = n_k \delta_{{\bf{kp}}},$$ 
where $n_k$ is the Bose distribution function. The correlators of the phonon fields for different spatial dimensions are
\begin{align}
\nonumber
&\text{1D:} \quad K_\text{1D}(x, t) = \frac{g^2 s}{4\pi r_c^2} \Biggl[
\frac{r_c^2}{(x+st)^2} - \frac{\pi^2}{\sinh^2\left(\frac{\pi(x+st)}{r_c}\right)} \\  
& \quad \quad \quad + \frac{r_c^2}{(x-st)^2} - \frac{\pi^2}{\sinh^2\left(\frac{\pi(x-st)}{r_c}\right)}
\Biggl], \\ 
&\text{2D:} \quad K_\text{2D}(x,t) = \frac{g^2}{2\pi} \int dk\ k \omega_k n_k \cos (\omega_k t) J_0 (k x), 
\label{K2D}
\\ 
&\text{3D:} \quad K_\text{3D}(x, t) = - \frac{1}{2\pi x} \frac{\partial K_{1D}(x,t)}{\partial x},
\end{align}
where 
\begin{align}
    \label{tauCPhon}
\tau_c = \frac{\hbar}{T_\text{ph}}, \quad r_c = s \tau_c,  
\end{align}
$x = |x_1-x_2|$ and $t= |t_1-t_2|$, and $x_i$ is the coordinate along the one-dimensional channel. In the case of 1D phonons, they propagate along the same direction as electrons. In general, phonons are characterized by the temperature $T_\text{ph}$ that may differ from the electronic temperature in the leads. This allows us to consider the zero-temperature electronic noise at finite phonon temperature (finite $\tau_c$).

\subsection{Ballistic regime: $K \tau_c^2 / \hbar^2 \gg 1$}
Here we summarize the expressions for dephasing rate in the ballistic regime for different phonon dimensionalities:
\begin{gather}
\text{1D:} \ \sqrt{\frac{\pi s g^2}{12 \hbar^2 r_c^2}}, \quad
\text{2D:} \ \sqrt{\frac{\zeta(3) s g^2}{2 \pi \hbar^2 r_c^3}}, \quad 
\text{3D:} \ \sqrt{\frac{\pi^2 s g^2}{ 60 \hbar^2 r_c^4}},
\end{gather}
where $\zeta(x)$ is the Riemann zeta function.

\subsection{Diffusive and logarithmic dephasing: $K \tau_c^2 / \hbar^2 \ll 1$}

\textit{1D case.} The phonon-induced dephasing rate is be calculated using Eq.~(\ref{tauvarphi}):
\begin{gather}
\label{D1Ddiff}
\frac{1}{\tau_\varphi} = \frac{g^2}{2 \hbar^2 r_c} = \frac{g^2}{2 \hbar^3 s} T_\text{ph}.
\end{gather}
In this calculation, we assume that phonons could transmit across barriers. The 1D dephasing rate grows linearly with the phonon temperature. The transmission coefficient is given by Eq.~(\ref{tDyn}). Interestingly, this dephasing rate has a peculiar limit of zero sound speed:
\begin{gather}
\frac{1}{\tau_\varphi} \sim \int d\tau K(\tau) \sim \frac{T}{s} \xrightarrow[s \rightarrow 0]{} \infty.	
\end{gather}
More accurately, there are two different cases:
\begin{align}
&\frac{s}{L} \gg T: \quad K(\tau) \sim \frac{T^2}{s} \e^{-T \tau},\quad \forall\, \tau,\\
&\frac{s}{L} \ll T: \quad K(\tau) \sim 
\begin{cases}
\dfrac{T}{L} \e^{-s\tau/L}, \quad \tau \ll \frac{L}{s},\\[0.5cm]
\dfrac{T^2}{s} \e^{-T\tau}, \quad \tau \gg \frac{L}{s}.
\end{cases}
\end{align}
Thus, if we start to decrease the sound speed to zero, eventually the second regime ($s/L \ll T$) is established. Then the correlator becomes time-independent for $Q \ll v/s$, which leads to
\begin{gather}
\av{\e^{i \int_{0}^{t}d\tau V(\tau)}} \approx \e^{- t^3 g^2 T s/L^2 }, \quad 
\frac{1}{\tau_\varphi} \xrightarrow[s \rightarrow 0]{} 0.
\end{gather}

\textit{2D case.} 
For phonons in the plane, the direct calculations of the dephasing rate by means of Eq.~(\ref{tauvarphi}) gives zero:  
\begin{gather}
\int_0^\infty d\tau K(\tau) = 0,
\end{gather}
as can be easily seen from the phonon correlator in two dimensions, Eq.~(\ref{K2D}).

In this case, we need to analyze more accurately the double integral in (\ref{expCalc}), which depends on the ratio between the structure size and the phonon correlation length. For short structures, one obtains
\begin{gather} 
\label{D2DLsmall}
\int_0^{t} \frac{d\tau}{\hbar} \int_0^{\tau} \frac{d\tau'}{\hbar}  K(\tau')  \approx \gamma_{2D} \ln\left(\frac{t}{\tau_c}\right), \quad \frac{L}{r_c} \ll 1, \\ \label{gammaDef}
\gamma_{2D} = \frac{g^2}{2 \pi \hbar^2 r_c s} = \frac{g^2}{2 \pi \hbar^3 s^2} T_\text{ph}. 
\end{gather}
For long structures, $L / r_c \gg 1$, at short times $t \ll L/s$, one gets a linear growth and for larger times a logarithmic grows, same as in Eq.~(\ref{D2DLsmall}):
\begin{gather} 
\label{D2DLbigg}
\int_0^{t} \frac{d\tau}{\hbar} \int_0^{\tau} \frac{d\tau'}{\hbar}  K(\tau') \approx 
\begin{cases}
\gamma_{2D} \frac{\pi s}{2 L} t, \quad &t \ll \frac{L}{s}, \\
\gamma_{2D} \ln\left(\frac{t}{\tau_c}\right), \quad &t \gg \frac{L}{s}.
\end{cases}
\end{gather}
This means that the random phase induced by the interaction with 2D phonons grows slower with time than for the diffusive phase dynamics. This is the reason why we get zero for the dephasing rate in Eq.~(\ref{expDiff}). For $\gamma \ll 1$, we can use the logarithmic result regardless of the relation between $L$ and $r_c$.
 
The typical value of electron-phonon coupling constant is $\lambda \sim 5 \cdot 10^{-11}$~erg, the material density is $\rho_{3D} = 5$~g/cm$^3$, and the surface phonon localization length $\kappa$ is an order of atomic length: $\kappa \sim 1$~nm$^{-1}$. With these values, one estimates $$\gamma_{2D}/T \sim 1 \cdot \text{K}^{-1}.$$

\textit{3D case.} In this case, we have the same scenario as for 2D phonons, but $\gamma_{3D}$ is now quadratic in the phonon temperature: 
\begin{gather} 
\label{gamma3D}
\gamma_{3D} = \frac{g^2}{2 \pi^2 \hbar^2 r_c^2 s} = \frac{g^2}{2 \pi^2 \hbar^4  s^3} T_\text{ph}^2.
\end{gather}
The limit of zero sound velocity for 2D and 3D phonons is taken in the same way as for 1D phonons. The conductance has a non-Breit-Wigner form. It is quite close to $e^2/h$, when the phonon temperature is the same as that for electrons in the contacts, since the characteristic dephasing-induced level broadening is of the same order as temperature. 
	 
\subsection{Fano factor for logarithmic dephasing}
As was shown above, for 2D and 3D phonons the logarithmic dephasing takes place. To calculate the Fano factor in this case, we assume that the leads are at zero temperature, while the phonons between the barriers are kept at non zero temperature $T_\text{ph}=T>0$. 
We then apply Eq.~(\ref{FanoLogExact}) for the Fano factor for logarithmic dephasing.
For example, in a system with the equal barriers, $\Gamma \equiv \Gamma_1 = \Gamma_2$, for 2D phonons at $\Gamma\ll eV \ll T$ we get from Eq.~(\ref{F-log}):
\begin{gather}
\label{FlogSimple}
F = \frac{1}{2} \left[ 1 - 
\frac{2}{\pi} \left(\frac{\Gamma}{T}\right)^{T g^2/\pi \hbar^3 s^2} \left(\frac{\Gamma}{eV}\right)^{1 - T g^2/\pi \hbar^3 s^2} 
\right].
\end{gather}

\section{Conclusion}

In this paper, we have addressed the problem of influence of inelastic processes on the Fano factor for a double-barrier structure. We have derived a general expression for the shot-noise power for a 1D conductor with arbitrary inelastic scattering. Next, we have considered the transmission of electrons through a double-barrier resonator with a random nonstationary potential inside it. We have derived exact inelastic  transmission and reflection amplitudes that depend on the random time-dependent potential. We have analyzed the transmission coefficient and the Fano factor for this structure, assuming two types of dynamics of electronic wave-function phases: diffusion type, $\langle \varphi_f^2 \rangle \sim t/\tau_\varphi$, and logarithmic, $\langle \varphi_f^2 \rangle \sim \ln \left( t/\tau_c \right)$. For the transmission coefficient, the diffusion type of dynamics leads to a Lorentzian shape with the width determined by a sum of elastic and dephasing contributions. The logarithmic type of dephasing leads to an unusual shape of the transmission coefficient as a function of energy, with the two scales given by the elastic width and the inverse correlation time of the fluctuating potential. The Fano factor for such structure also depends of dephasing type. For the diffusion type of dephasing, the Fano factor is largely insensitive to dephasing up to small corrections at low bias voltage. For the logarithmic type, there is a strong dependence of the Fano factor on dephasing rate at low voltages, when the bias is within a single broadened level. Finally, we have applied our general formalism to the case when the fluctuating potential is produced by phonons. For 1D phonons, the dephasing is of diffusion type with the rate proportional to the phonon temperature. For 2D and 3D phonons, we have found that dephasing is of logarithmic type, which leads to an unusual temperature dependence  of the Fano factor.
To conclude, a quantum-mechanical derivation of shot noise through inelastic transmission and reflection amplitudes and averaging over random potential reveals a dependence of the Fano factor on the type of wave function phase dynamics.

\begin{acknowledgments}
The authors thank I.V. Gornyi for his great contribution to this work. 
This work was supported by the Russian
Foundation for Basic Research (Grant No. 22-12-00139).
\end{acknowledgments}

\appendix

\section{Example: transmission across time-dependent delta barrier}
\label{Appendix1}

We start from simple example - obtain transmission amplitude across scattering from time dependent delta barrier, assuming that this time dependence is random but well defined function in time. The Schrodinger equation is:
\begin{gather}
    -\frac{\hbar^2}{2m}\frac{\partial^2 \psi}{\partial x^2} + f(t) \delta(x) \psi = i \hbar \frac{\partial \psi}{\partial t}.
\end{gather}
Defining wave function in basis of incident waves we can write it in next form:
\begin{align}
    \psi_< &= \e^{ikx-iE_kt/\hbar} + \int dp\, r(p,k) \e^{-ipx - iE_p t/\hbar}, \quad &x < 0,\\
    \psi_> &= \int dp\, t(p,k) \e^{ipx - iE_p t/\hbar}, \quad &x > 0,
\end{align}
where we assume that $k$ wave incident from the left side and all energies are contribute to transmission and backscattering. 
From Schrodinger equation we can find conditions at the delta function position:
\begin{align}
    \psi_<(0) &= \psi_>(0), \\
    \frac{2m}{\hbar^2} f(t) \psi_<(0) &= \frac{\partial \psi_>}{\partial x}(0) - 
    \frac{\partial \psi_<}{\partial x}(0).
\end{align}
This conditions should be valid for any time and we can obtain the system of equations by integrating with $\e^{iE_q t/\hbar}$. Next we use one approximation which will allow us to solve task exactly - linearising spectra $E_k \approx v \hbar k$. With this assumptions we can find next equations definig transmission and reflection amplitudes:
\begin{align}
    \delta(q-k) + r(q,k) &= t(q,k), \\ \label{deltaFEq}
    \int dp\, \frac{2mv}{\hbar^2} f(q-p) t(p,k) &= i q [t(q,k) -\delta(q-k) + r(q,k)], \\
    f(q-p) &= \int dt \, f(t) \e^{iv(q-p)t}.
\end{align}
Solving this equations we obtain transmission amplitude throw random but well time defined function $f(t)$:
\begin{gather}
    t(p,k) = -ik \int dx\int^x dy\, \e^{ipx-iky} 
    \exp \left( - \int_y^x dz \frac{m}{\hbar^2} f\left( \frac{z}{v}\right)   \right).
\end{gather}
If particles energy changes is small due to time dependent potential $|p-k|/k \ll 1$ we can replace prefactor in right side of eq.~(\ref{deltaFEq}) on $q \rightarrow k$ and obtain next equation for transmission amplitude:
\begin{gather}
    t(p,k) = \int dx\, \frac{i k}{ik - f(x/v) /\hbar}\e^{i(p-k)x}.
\end{gather}
For the case of stationary barrier $f(t) = \lambda$ this leads to well known result:
\begin{gather}
    t(p,k) = \frac{i k}{ik - \lambda /\hbar} \delta(p-k).
\end{gather}

\section{Transmission and reflection amplitude for a double-barrier structure with dephasing}
\label{Appendix2}

In this Appendix, we obtain transmission coefficient throw the double barrier structure with random non stationary potential. As we mentioned in main text for the weak and smooth potential wave function take additional phase only, see Eq.~(\ref{psiDep}). 

For system with barriers the general wave function moving from the left with incident wave vector $k$ is:
\begin{align*}
    \psi_I &= \e^{ik(x-vt)} + \sum_p r_{pk} \e^{-ip(x+vt)},\\
    \psi_{II} &= \alpha(x,t) \sum_p a_{pk} \e^{ip(x-vt)} + \beta(x,t) \sum_p b_{pk} \e^{-ip(x+vt)},\\
    \psi_{III} &= \sum_p t_{pk} \e^{ip(x-vt)}, \\
    \alpha(x,t) &= \exp\left(-i\int^t d\tau V(x+vt-v\tau,\tau)\right), \\
    \beta(x,t) &= \exp\left(-i\int^t d\tau V(x-vt+v\tau,\tau)\right).
\end{align*}
Here region I, II and III corresponds to regions before, between and after double barriers. In order to obtain transmission and reflection coefficients we need to use scattering matrices of barriers linking ingoing and outgoing waves. At this step we can assume that barriers are point like and add some extra phase to reflection and transmission amplitudes without loss of generality. Barriers located at the $0$ and $L$ points. The condition on left barrier is:
\begin{align}
\label{sMatrEq}
    \left(\begin{matrix}
    t_1 && r_1 \\
    r_1 && t_1
    \end{matrix} \right)
    \left(\begin{matrix}
    \psi_I^> \\
    \psi_{II}^<
    \end{matrix} \right) = 
    \left(\begin{matrix}
    \psi_{II}^> \\
    \psi_{I}^<
    \end{matrix} \right),
\end{align}
where index $>(<)$ correspond to the right (left) moving parts of wave function. We neglect dependence of barriers transmission/reflection amplitude on wave vector, assuming that random potential weak and transmitted particles wave vector changes is small $|p-k|/k \ll 1$. So we can take it on incident wave vector, i.e. $t_i = t_i(k)$. From the equation (\ref{sMatrEq}) and the same one on second  barrier we can obtain:
\begin{align}
\nonumber
   &\frac{\alpha(0,t)}{t_1} \sum_p a_{pk} \e^{-ipvt} - \frac{r_1 \beta(0,t)}{t_1} \sum_p b_{pk} \e^{-ipvt} = \e^{-ikvt}, \\
\nonumber  
   &\sum_p t_{pk} \e^{ip(L-vt)} = t_2 \alpha(L,t) \sum_p a_{pk} \e^{ip(L-vt)}, \\ \nonumber
   &r_2 \alpha(L,t) \sum_p a_{pk} \e^{ip(L-vt)} = \beta(L,t) \sum_p b_{pk} \e^{-ip(L+vt)}.
\end{align}
Using this equations one can obtain transmission coefficient:
\begin{align}
 &t_{pk} = \int dt\, v\e^{i pvt} t_1t_2 \left( 1 - r_1r_2 F(t) \e^{-\tau_f \frac{\partial}{\partial t}} \right)^{-1} 
 G(t) \e^{-ikvt}, \\
 & F(t) = \frac{\alpha(L,t + \tau_f/2) \beta(0,t)}{\alpha(0,t) \beta(L,t - \tau_f/2)}, \quad G(t) = \frac{\alpha(L,t + \tau_f/2)}{\alpha(0,t)}.
\end{align}
By expanding this equation in series on reflections between barriers, assuming that random potential varying slowly on time flight scale $\partial V(x,t) / \partial t \ll V(x,t) / \tau_f$:
\begin{gather} \nonumber
    t_{pk} =  \int dx\,  \e^{i(k-p)x}  t_1t_2 \sum_{n=0}^\infty \left( r_1 r_2 \e^{i2kL} \right)^{n}  \\
     \cdot \exp \left(
    -i \int_{-\tau_f n}^0 d\tau\, V(x_{in}(\tau), \tau - x/v)
    \right).
\label{tPKexact}
\end{gather}
$x_{in}(t)$ - trajectory of particle between barriers, describing reflecting motion.

Using same method we can obtain reflection amplitude:
\begin{gather}
    r_{pk} = \int dtdq\, v\e^{i (p-q)vt} 
    \frac{r_2 F(t) \e^{i2qL} - r_1^*}{G(t) t^*_1 t_2} t_{qk}.
\end{gather}
For small potential ($K \tau_c^2 \ll 1$), $F(t) \approx G(t) \approx 1$ and we obtain simple relation between reflection and transmission amplitudes:
\begin{gather}
    r_{pk} = \frac{r_2 \e^{ipL} -r_1^*\e^{-ipL} }{\e^{-ipL}t_1^* t_2} t_{pk}.
\end{gather}

\bibliography{main}
 
\end{document}